\documentclass[aip,reprint]{revtex4-2}

\newcommand{\red}[1]{{\color{black}#1}}

\usepackage{xcolor,textcomp}
\usepackage{graphicx,hyperref,epstopdf,mathtools}
\usepackage{amsmath,amssymb,multirow,array,gensymb}
\usepackage{fancyhdr,float,accents}
\usepackage[none]{hyphenat}
\usepackage{enumitem,relsize}
\usepackage{footmisc}
\usepackage{soul}
\usepackage{tikz}
\usepackage{amsbsy,mathrsfs}
\usepackage{tabularx,booktabs}

\begin{document}

\preprint{AIP/123-QED}

\title{Interfacial Transition Waves and Domain Wall Self-Pinning in Bi-stable Mechanical Metamaterial}

\author{Michael J. Frazier}
\email[Corresponding author email: ]{mjfrazier@ucsd.edu}
\affiliation{Department of Mechanical and Aerospace Engineering, University of California, San Diego, La Jolla, California 92093, USA}
\affiliation{Program in Materials Science and Engineering, University of California, San Diego, La Jolla, California 92093, USA}


\begin{abstract}
\red{The understanding of transition waves in mechanical metamaterials currently derives from architectures fitting one of three archetypes as distinguished by the stability and order parameter: multi-stable with scalar or vector order parameter, and bi-stable with scalar order parameter.}
\red{However, a fourth archetype is conspicuously absent: bi-stable with vector order parameter.}
\red{Completing the tetrad, in this article,} we numerically examine the \red{transformation} dynamics \red{unique to bi-stable mechanical metamaterials with vector order parameter.}
\red{We report two types of} interfacial transition waves, \red{whose propagation is restricted} to the narrow region of \red{configurational boundaries (e.g., domain walls).}
\red{We show that the wave speed may depend on its polarization.}
\red{In addition, we reveal an approach to regulate the severity of discreteness effects, which elevates} discreteness as an intrinsic, tunable mechanism \red{for} controlling the wave speed -- \red{including pinning/de-pinning of the domain wall} -- without altering the \red{metamaterial architecture}.
\red{The results support applications in, e.g., reconfigurable mechanical circuits and complex domain patterning}, and motivate further investigation of metamaterials with unconventional transition processes.
\end{abstract}

\maketitle

\section{Introduction}
Structural phase transformation\cite{Wadhawan2000,Fujimoto2005} is an energy minimizing process occurring in solids (e.g., ferroics) whereby
the material microstructure adjusts from one equilibrium configuration
to another as differentiated by an order parameter.
The spatially non-uniform process produces regions of homogeneous configuration (i.e., domains) separated by an interpolating interface (i.e., domain wall) that, in propagating, constitutes a transition wave and locally alters the microstructure, facilitating the transformation.
The ability to create, assess, and reliably manipulate transition waves is \red{essential} to proposed nanotechnologies\cite{CatalanRMP2012,SharmaM2019,MeierNRM2022};
\red{although, the requisite} accessibility of the microstructure is \red{complicated} by the disparity in the relevant spatio-temporal scales.


Mechanical metamaterials comprising multi-stable structural units demonstrate similar transformation phenomena\cite{NadkarniPRL2016,ShirokyJMPS2017,FrazierAM2017,KhajehtourianEML2022} at an experimentally accessible, structural level. 
The ability to customize the internal architecture for bespoke transformation environments grants exceptional command of the metamaterial response and facilitates the creation of functional designs.
To develop the fundamental theoretical understanding,
prior studies have, e.g., analyzed the \red{transformation} response of exotic architectures\cite{HwangPRE2018,WangEML2023,VeenstraNature2024}, examined the effect of discreteness\cite{TangARXIV}, and \red{revealed universal} scaling laws\cite{NadkarniPRB2016,RayPRM2025}.
To date, the proposed applications span, e.g., mechanical memory\cite{PechacAPL2023} and computing\cite{RaneyPNAS2016,PalPNAS2023,ByunNC2024}, soft robot locomotion\cite{DengPRAppl2022}, reconfigurable\cite{JanbazNComm2024} and deployable structures\cite{ZareeiPNAS2020}, and elastic wave control\cite{RamakrishnanJAP2020,WatkinsPRB2021}.



\red{From a review of the metamaterial literature, it is apparent that, despite the diversity of architectures, the current understanding of transition wave dynamics has been garnered from only three metamaterial archetypes: multi-stable systems characterized by scalar\cite{WangAPL2023,RayPRM2025} and vector\cite{KhajehtourianEML2022,ZhouCP2024,ZhouCP2024_Polar} order parameters, and bi-stable systems described by a scalar order parameter\cite{NadkarniPRE2014}.}
\red{Each archetype has an associated energy landscape whose peculiarities extend to the attributes of the supported transition waves.}
\red{Of their order parameters, scalar transition waves require only change in magnitude while vector waves require magnitudes and orientations change simultaneously, which raises the notion of wave polarization.}
\red{However, a fourth archetype is conspicuously absent from a supposed tetrad: bi-stable systems characterized by a vector order parameter.}
\red{In this case, the wave polarization would be a transformable quality like the system configuration.}


\begin{figure*}[t]
    \centering
    \includegraphics{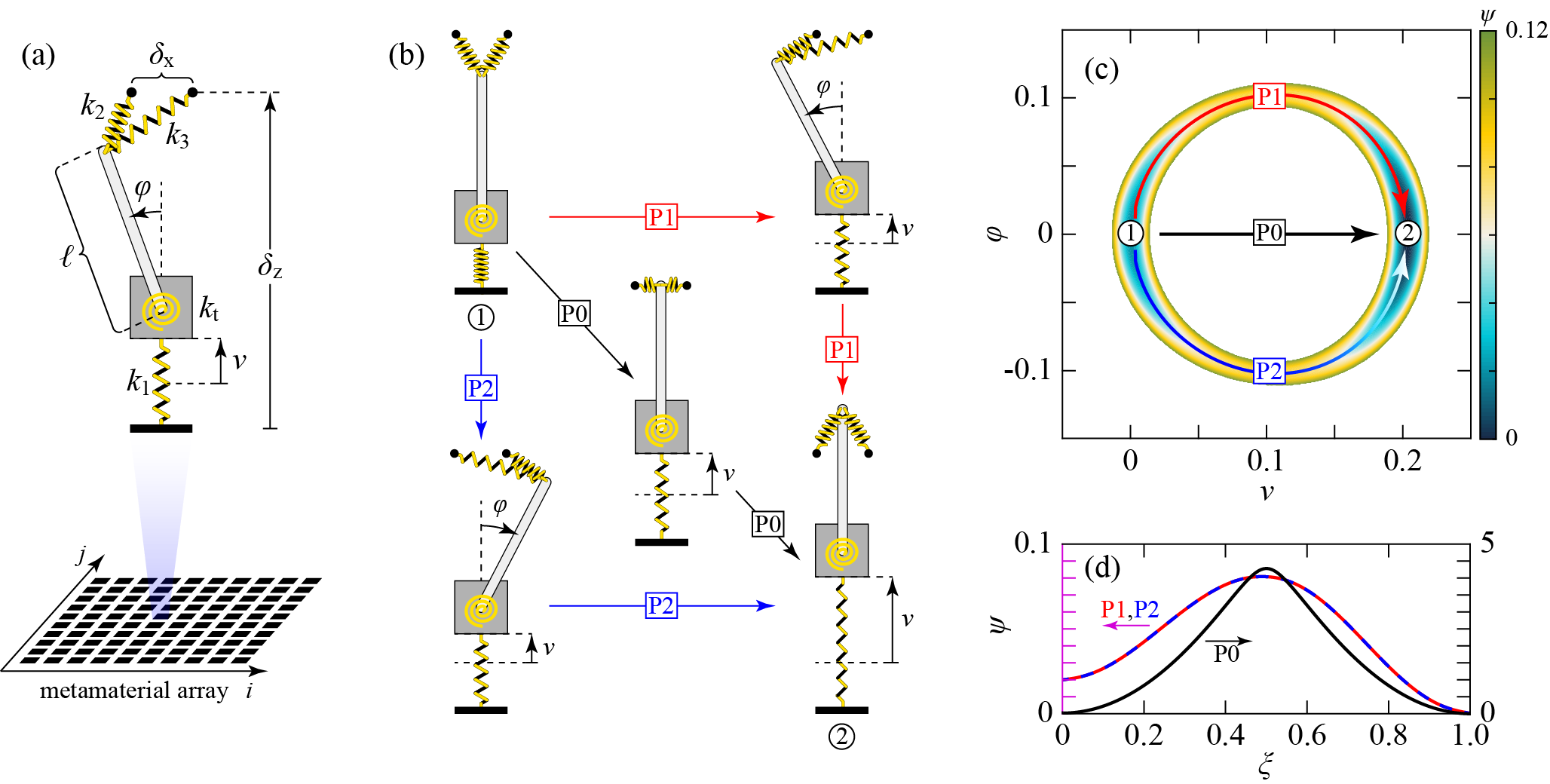}
	\caption{\red{Multi}-path Transformation.
    (a) Unit cell of representative \red{bi-stable} mechanical metamaterial.
    \red{(b) Schematic illustrating three sequences of unit cell deformations to effect a transformation between states $\mathbf{s}_1$ and $\mathbf{s}_2$: P1 and P2 are minimum energy pathways; P0 is a high energy pathway.}
    (c) Map of the local potential landscape, $\psi(v,\varphi)$, showcasing two local minima, $\mathbf{s}_1$ and $\mathbf{s}_2$, connected by two unique low-barrier transition pathways (i.e., polarizations), P1 and P2 [for clarity, $\psi(v,\varphi)>0.1$ regions clipped].
    (d) The potential energy, \red{$\psi(\xi)=\psi[v(\xi),\varphi(\xi)]$}, where \red{$\xi\in[0,1]$ parametrizes the} low-barrier paths, P1 (red, dashed) and P2 (blue, solid), and the high-\red{barrier} direct path, P0 (black, solid).
    Parameter values: $\alpha=10/11$, $c=1/5$, $k_1=10^{-3}$, $k_2=1/2$, $k_\mathrm{t}=5\times10^{-3}$, $k=1/4$, $\ell=1$, $\delta_x=1/100$, $\delta_\mathrm{z}=551/500$, $v_0=1/5$, $\varphi_0=0$.}
	\label{fig:Model}
\end{figure*}

\begin{figure*}[t]
    \centering
    \includegraphics{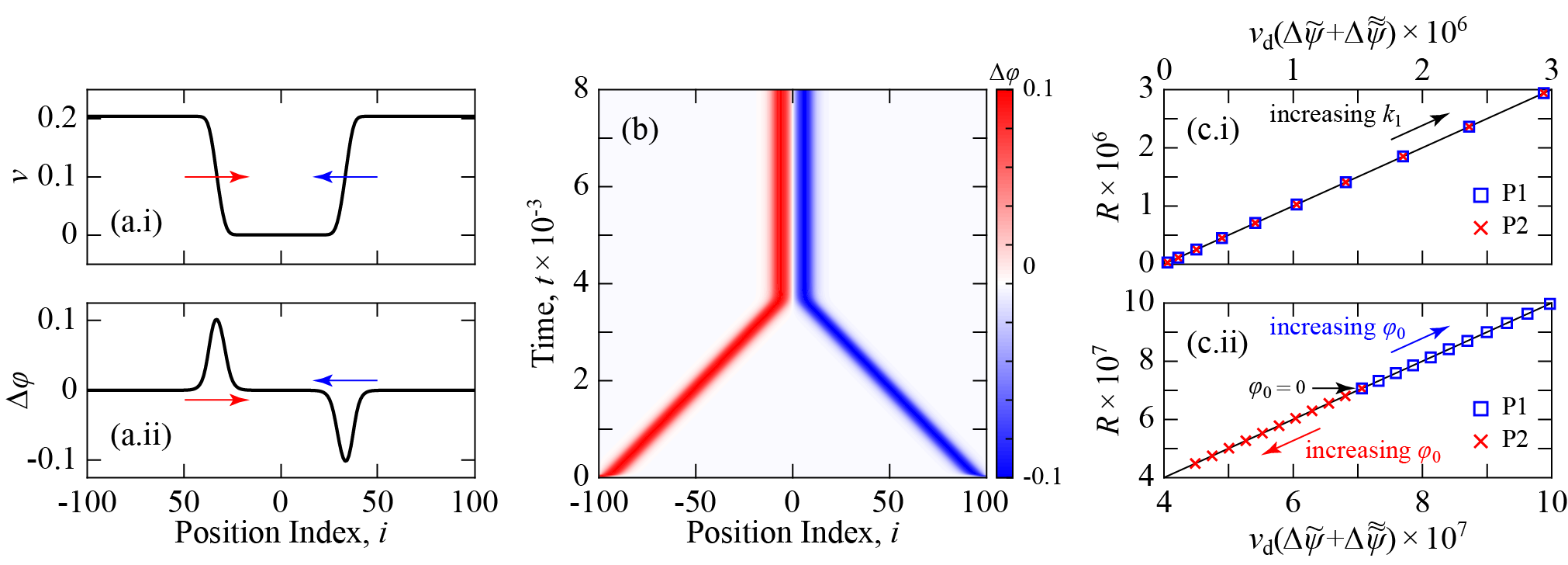}
	\caption{\red{Multi}-path Phase Transitions.
    (a) Profiles of two plane transition waves of opposing polarization propagating toward each other (arrows indicate propagation direction).
    (b) Spatiotemporal configuration diagram depicting the propagation and collision of two planar transition waves of opposing polarization.
    \red{(c) Validation of the scaling law in Eq. \eqref{eq:USL} for transition waves under (i) different potential driving effected by varying $k_1$ and (ii) path-dependent energy landscapes achieved through varying $\varphi_0$.}
    \red{Parameter values: (a,b) identical to Fig. \ref{fig:Model} caption; (c) identical to Fig. \ref{fig:Model} caption, except (i) $k_1=2p\times10^{-4}$ and (ii) $\varphi_0=p/100$, where $p=0,\ldots,10$.}}
	\label{fig:USL}
\end{figure*}

In this article, we introduce a bi-stable metamaterial model \red{for which the order parameter is, nonetheless, a vector quantity and there is more than one minimum energy pathway connecting stable states in the energy landscape, engendering transition waves with a polarization.}
This novel condition impacts the transformation behavior to yield several results unique to the metamaterial literature and which may not otherwise manifest, including (i) interfacial transition waves confined to configurational boundaries, (ii) polarization-dependent domain wall speed, (iii) domain walls expressing multiple polarizations simultaneously, and (iv) 2D domain morphologies stabilized by custom, reversible domain wall self-pinning.
Specifically, the premier and latter results support the notions of reconfigurable mechanical circuits and custom domain patterning, respectively.
In general, these results motivate further exploration of metamaterials with unconventional energy landscapes.

\section{Model}
Figure \ref{fig:Model}a depicts the fundamental unit of the 2D phase-transforming metamaterial considered herein.
In particular, the unit cell comprises a mass restricted to vertical displacement, $v$, to which a rigid rod with additional in-plane, angular displacement, $\varphi\in[-\pi,\pi]$, is pin-connected.
Penalizing the displacements \red{is} a set of elastic elements, including axial springs, $k_i$, $i=1,\ldots,3$ that connect the mass and rod to fixed supports, and torsional spring, $\kappa_\varphi$, at the pin connection.
Thus, the local potential energy function has the form\cite{SI}:
\begin{align*}
	\psi(v,\phi)&=\frac{1}{2}\kappa_\varphi(\varphi-\varphi_0)^2+\frac{1}{2}k_1(v-v_0)^2\cdots\\
	&+\frac{1}{2}k_2\left[\ell_2(v,\varphi)-\ell_{20}\right]^2+\frac{1}{2}k_3\left[\ell_3(v,\varphi)-\ell_{30}\right]^2,
\end{align*}
where $\varphi_0$ and $v_0$ are pre-displacements, and $\ell_i(v,\varphi)$ and $\ell_{i0}$ are, respectively, the instantaneous and relaxed lengths of springs, $k_i$.
Apparently, $\ell_1(v,\varphi)=v-v_0+\ell_{10}$. In the following, for simplicity, $k_3=k_2$.

For certain values of the \red{unit cell} geometric and elastic parameters, $\psi(v,\varphi)$ realizes two \red{local minima} at $\mathbf{s}_i:=(v_{\mathbf{s}i},\varphi_{\mathbf{s}i})$, \red{$i=1,2$} indicative of the bi-stability of the unit cell configuration, a condition \red{likewise achieved through various alternative unit cell} designs in the metamaterial literature.
\red{However, different from previous manifestations of bi-stability, the current local minima are joined by more than one minimum energy pathway; thus, endowing state transitions with a polarization.}
\red{Typically, the polarization results from symmetry breaking via
the unit cell construction alone; however, here, the polarization stems from symmetry breaking provoked by an internal variable.}
\red{Specifically, symmetry breaking originating from the unit cell construction manifests the two stable states differentiated only by $v$, which, contrary to convention, is also the relevant control variable provoking an additional symmetry breaking at critical values, which results in two stable states of $\varphi(v)$ consistent with the two minimum energy pathways and representative of two polarizations as distinguished by the sign of $\Delta\varphi=\varphi-\varphi_{\mathbf{s}i}$.}
\red{In the following, we explore the transformation behavior uniquely supported by the bi-stable potential with multiple transition pathways that give rise to polarization.}

\red{The unusual energy landscape of the unit cell is reminiscent of that associated with the stacking configuration of weakly interacting atomic layers in sliding ferroelectrics\cite{LiACS2017,SternScience2021,WuPNAS2021}.}
\red{At the structural level, amenable to metamaterial fabrication, a qualitatively} similar energy landscape \red{emerges from}, e.g., a unit cell comprising a buckled beam\cite{RadissonPRL2012} or membrane; however, the simplicity of the present lumped-parameter model \red{helps clarify} the structure-response relation.



The metamaterial emerges from a 2D array of the above-described unit cells with linear elastic, nearest-neighbor interactions that penalize differences in configuration among the coupled cells.
For a square lattice with spacing, $a$, the dimensionless governing equations of the system are\cite{SI}:
\begin{subequations} \label{eq:disc_EOM}
    \begin{align}
        \ddot{v}_{i,j}&+c\dot{v}_{i,j}-\frac{\alpha}{2}\left[\cos(\varphi_{i,j})\dot{\varphi}_{i,j}^2+\sin(\varphi_{i,j})\ddot{\varphi}_{i,j}\right]\cdots\nonumber\\
        &+k_v(2v_{i,j}-v_{i+1,j}-v_{i-1,j})\cdots\nonumber\\
        &+k_v(2v_{i,j}-v_{i,j+1}-v_{i,j-1})\cdots\nonumber\\
        &+\tilde{\psi},_{v}(v_{i,j},\varphi_{i,j})=0,\\
        \ddot{\varphi}_{i,j}&+c\dot{\varphi}_{i,j}-2\sin(\varphi_{i,j})\ddot{v}_{i,j}\cdots\nonumber\\
        &+k_\varphi(2\varphi_{i,j}-\varphi_{i+1,j}-\varphi_{i-1,j})\cdots\nonumber\\
        &+k_\varphi(2\varphi_{i,j}-\varphi_{i,j+1}-\varphi_{i,j-1})\cdots\nonumber\\
        &+\tilde{\tilde{\psi}},_\varphi(v_{i,j},\varphi_{i,j})=0,
    \end{align}
\end{subequations}
where
$\alpha=m_\mathrm{r}/m_\mathrm{tot}$ relates the rod and unit cell masses, $m_\mathrm{r}$ and $m_\mathrm{tot}$, respectively, $c$ is the on-site viscosity accounting for losses, $k_v$ and $k_\varphi$ are the inter-site coupling strengths, and $\{i,j\}$ are the unit cell position indices.
For much of the article, $k_v=k_\varphi=k$.
In addition, the \red{chosen} geometric parameters ensure $|\varphi_{i,j}|\ll1$ and $\dot{\varphi}_{i,j}\ll1$, \red{causing the inertial coupling to effectively vanish; thus, Eqs. \eqref{eq:disc_EOM} simplify to:}
\begin{align} \label{eq:disc_EOM_uncpld}
    \red{\ddot{\mathbf{u}}_{i,j}+c\dot{\mathbf{u}}_{i,j}}&\red{+k(2\mathbf{u}_{i,j}-\mathbf{u}_{i+1,j}-\mathbf{u}_{i-1,j})\cdots}\nonumber\\
    &\red{+k(2\mathbf{u}_{i,j}-\mathbf{u}_{i,j+1}-\mathbf{u}_{i,j-1})\cdots}\nonumber\\
    &\red{+\psi,_\mathbf{u}(v_{i,j},\varphi_{i,j})=\mathbf{0},}
\end{align}
where \red{$\mathbf{u}_{i,j}^\mathrm{T}=[v_{i,j},\varphi_{i,j}]$} is the relevant vector order parameter \red{and $\psi,_\mathbf{u}^\mathrm{T}=[\tilde{\psi},_v,\tilde{\tilde{\psi}},_\varphi]$}.
A 2nd-order Taylor expansion of Eqs. \eqref{eq:disc_EOM_uncpld} yields a corresponding continuum approximation consistent with sufficiently smooth variation in \red{$\mathbf{u}_{i,j}$}:
\begin{equation} \label{eq:continuum_EOM}
    \ddot{\mathbf{u}}+c\dot{\mathbf{u}}-k\nabla^2{\mathbf{u}}+\psi,_{\mathbf{u}}(v,\varphi)=\mathbf{0}.
\end{equation}
In the following, the simulated metamaterial response stems from numerical integration of the full discrete model [Eqs. \eqref{eq:disc_EOM}] while the analysis proceeds from the simplified continuum model [Eq. \eqref{eq:continuum_EOM}].

\section{Results}
\subsection{Polarized Transition Waves}
We investigate the propagation of plane transition waves within a metamaterial of the unit cells described in Fig. \ref{fig:Model}, specifically, a $201\times201$ array with periodic boundary conditions in the $j$-dimension.
Accordingly, the associated on-site potential (Fig. \ref{fig:Model}c) possesses local minima at $\mathbf{s}_i:=(v_{\mathbf{s}i},\varphi_{\mathbf{s}i})$, $i=1,2$, connected by low-barrier transition pathways, P1 \red{($\Delta\varphi\geq0\;\forall{v}$)} and P2 \red{($\Delta\varphi\leq0\;\forall{v}$)}, respectively.
The energy barrier along P0 \red{($\Delta\varphi=0\;\forall{v}$)} is prohibitive; approximately forty times that of either P1 or P2 (Fig. \ref{fig:Model}d).

\begin{figure*}[t]
    \centering
    \includegraphics{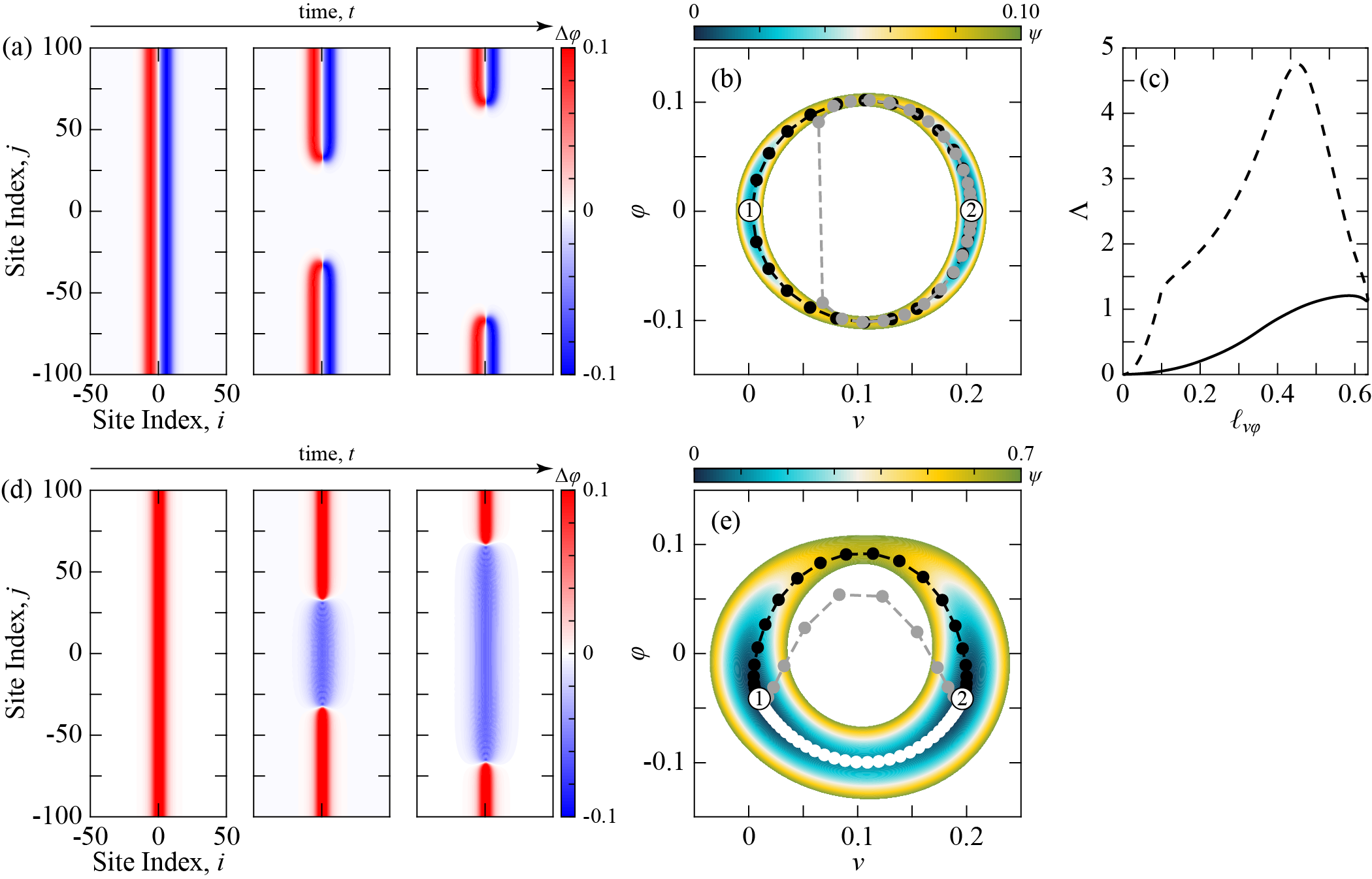}
	\caption{Interfacial Transition Waves.
    (a,d) Snapshots depicting the evolution of (a) the interface-eliminating and (d) the polarization-switching interfacial transition waves.
    (b,e) Map of $\psi(v,\varphi)$ overlaid with the configurations of the 51 lattice sites (dots) along $j=50$ cross-section and nearest (b) the polarization boundary and (e) the center of a P1 domain wall.
    Shading from black to white indicates advancing time.
    The interfacial transition waves are shown to effectuate the transformation (b) by sending all $\mathbf{u}_{i,j}$ to $\mathbf{s}_2$ via P1 and P2 and, thus, eliminating the interface, and (e) by switching the transition pathway.
    (c) The total potential energy, $\Lambda$, of lattice sites along the $j=50$ cross-section for an interface-eliminating transformation achieved exclusively via the low-barrier pathways (solid) as well as the case where the configuration of the unit cell at $i=0$ is made to cross the central energy barrier toward $\mathbf{s}_2$ (dashed).
    $\ell_{v\varphi}$ is the length of the curve defined by $(v_{i,50},\varphi_{i,50})$ in configuration space.
    Parameter values: (a--c) identical to Fig. \ref{fig:Model} caption; (d,e) identical to Fig. \ref{fig:Model} caption, except $k=2$, $k_1=0$, $k_2=1/20$, $\varphi_0=-1/20$.}
	\label{fig:Transformations}
\end{figure*}

The metamaterial is initially uniform in the meta-stable state, $\mathbf{s}_1$; however, at the opposing boundaries we provoke \red{a} transformation to the ground state, $\mathbf{s}_2$, by \red{stimulating counter-propagating} transition wavefronts along P1 and P2, respectively.
Subsequently, the potential difference, $\Delta\psi=\psi(\mathbf{s}_1)-\psi(\mathbf{s}_2)>0$, drives the ensuing domain wall motion toward the system interior.
Figure \ref{fig:USL}a displays a snapshot of the displacement components along the $i$-dimension, revealing the unique profiles of \red{the polarized} wavefronts: \red{$\Delta\varphi$ distinguishes one polarization from the other}.
Figure \ref{fig:USL}b adds a temporal dimension, showing the \red{differently polarized} domain walls to propagate
until, upon collision, rather than annihilate, they form an interface analogous to a domain wall, except that the boundary interpolates \red{the distinct} polarizations rather than stable states (MOV. S1); consequently, the interface is not topologically protected.
\red{On the contrary, the} interface \red{resulting from} the collision of transition waves \red{transforming the system from a common} initial state \red{into} different final states -- \red{as occurs in multi-stable systems\cite{ZhouCP2024,ZhouCP2024_Polar}} -- is protected \red{by the opposing states}.

\red{Scaling laws\cite{NadkarniPRB2016,KhajehtourianEML2022,RayPRM2025} analytically describe the connection between the observed transformation dynamics and properties of the supporting medium.}
\red{For planar wavefronts with smooth variation in $\mathbf{u}_{i,j}$, Eq. \eqref{eq:continuum_EOM} reduces to a 1D form with solution, $\mathbf{u}(\mathbf{x},t)=\mathbf{u}(\mathbf{x}\cdot\hat{\mathbf{e}}-v_\mathrm{d}t)=\mathbf{u}(\sigma)$, where $\mathbf{x}$ is the position, $\hat{\mathbf{e}}$ is the propagation direction, $v_\mathrm{d}$ is the wave speed, and $\sigma$ is the wave coordinate.}
\red{Accordingly, Eq. \eqref{eq:continuum_EOM} becomes:}
\begin{equation*} 
    \red{(v_\mathrm{d}^2-k)\mathbf{u},_{\sigma\sigma}-v_\mathrm{d}c\mathbf{u},_\sigma+\psi,_{\mathbf{u}}(v,\varphi)=\mathbf{0}.}
\end{equation*}
\red{Pre-multiplying by $-v_\mathrm{d}\mathbf{u},_\sigma^\mathrm{T}$ and integrating over the domain $\sigma\in[-\infty,\infty]$ yields:}
\begin{equation} \label{eq:USL}
    \red{R=v_\mathrm{d}^2c\int_{-\infty}^{\infty}|\mathbf{u},_\sigma|^2\mathrm{d}\sigma=v_\mathrm{d}(\Delta\tilde{\psi}+\Delta\tilde{\tilde{\psi}}),}
\end{equation}
\red{since the first integral term vanishes far from the domain boundary.}
\red{Equation \eqref{eq:USL} describes a linear connection between the total power dissipated by the domain wall, $R$, and the potential driving furnished by the metamaterial, $\Delta\tilde{\psi}+\Delta\tilde{\tilde{\psi}}$.}

\red{Figure \ref{fig:USL}c.i plots $R$ versus $v_\mathrm{d}(\Delta\tilde{\psi}+\Delta\tilde{\tilde{\psi}})$ for plane transition waves under various potential driving as mediated by $k_1$.}
\red{The results lie along a line of slope 1, demonstrating the validity of the scaling law.}
\red{For $\varphi_0=0$, the dynamics of the oppositely polarized} wavefronts \red{are} identical; however, setting $\varphi_0\neq0$ \red{causes} the \red{dynamics to} diverge (Figs. \ref{fig:USL}c.ii and S1).
\red{While bi-stability restricts transition waves of opposing polarization to the same potential driving, the specific variation in the on-site energy along a particular pathway affects the profile of the corresponding transition wave, which is reflected quantitatively through $|\mathbf{u},_\sigma|^2$.}
\red{For $\varphi_0\neq0$, the energy landscape is no longer symmetric about $\varphi=\varphi_{\mathrm{s}i}$; therefore, transformations via P1 and P2 are not energetically equivalent, and the dynamics of transition waves become polarization-dependent.}
\red{Figure \ref{fig:USL}c.ii plots $R$ versus $v_\mathrm{d}(\Delta\tilde{\psi}+\Delta\tilde{\tilde{\psi}})$ for various (a)symmetric on-site potential as mediated by $\varphi_0\geq0$.}
\red{Apparently, the validity of the scaling law remains intact.}




\subsection{Interfacial Transition Waves}
The presence of the polarization interface represents a meta-stable state of the whole metamaterial array; however, as \red{this configurational} boundary is not topologically protected, it is yet possible to transform the metamaterial with such boundary into one without it.
To demonstrate, upon the lattice sites of \red{the} cross-section \red{defined by} $j=0$, we impose a quasi-static, linear transformation of the \red{current unit cell} configurations \red{comprising} the polarization interface to \red{configurations uniformly} in $\mathbf{s}_2$, which stimulates two, counter-propagating transition waves driven by the potential difference between $j$ cross-sections with and without the interface (MOV. S2).
In this scenario, wave propagation entails eliminating the interface.
Through a series of snapshots, Fig. \ref{fig:Transformations}a depicts the motion of the two interface-eliminating \red{wavefronts}.
Figure \ref{fig:Transformations}b plots the trajectories in configuration space of the lattice sites composing the $j=50$ cross-section as the interface-eliminating wave passes, providing \red{microstructure-level insight into the transformation process}.
Apparently, the transition wave propagates as the unit cell configurations transform to $\mathbf{s}_2$ via the low-barrier pathways exclusively (MOV. S2); although, for \red{$\psi(v,\varphi)$} with sufficiently low central energy barrier, transitions outside the low-barrier pathways is also possible (not shown).
Figure \ref{fig:Transformations}c compares the total potential energy, \red{$\Lambda=\sum_i\frac{1}{2}k(\mathbf{u}_{i,j}-\mathbf{u}_{i+1,j})^2+\psi,_{\mathbf{u}}(v_{i,j},\varphi_{i,j})$, $j=50$}, of a cross-section where the transformation is achieved exclusively via the low-barrier pathways and one where the configuration of the unit cell nearest the center of the polarization boundary is prescribed the direct path, P0, that crosses the central energy barrier, showing the former to be the more energetically favorable scenario.

\begin{figure*}[t]
    \centering
    \includegraphics{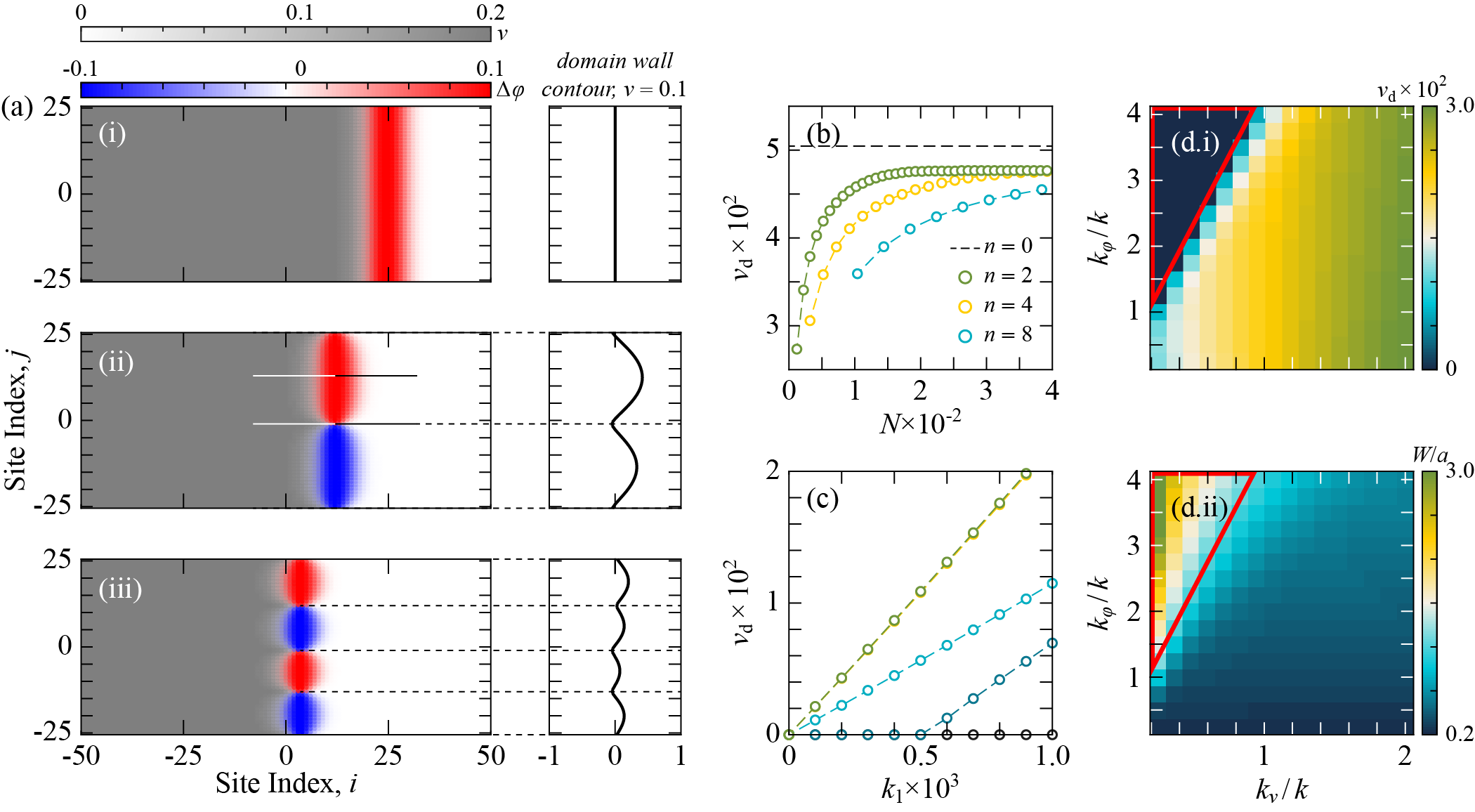}
	\caption{\red{Multi}-Polarized Domain Wall Motion.
    (a) Snapshots depicting the position of mobile domain walls of varying complexity following equal propagation time (arrows indicate propagation direction).
    Corresponding outset plots display the contour of level $v=0.1$, revealing the domain wall distortion as cross-sections of the wavefront nearer switch points lag those at greater distance.
    (b) Speed of domain wall with $n$ evenly distributed switch points \red{along} the $j$ dimension of a $N\times101$ metamaterial array.
    \red{Switch points diminish the wave speed; although, decreasing the density of switch points mitigates this effect.}
    \red{(c) Speed of multi-polarized domain wall ($n=2$) under variable driving (mediated by $k_1$) for different inter-site coupling, $(k_v/k,k_\varphi/k)$: results for $(1,1)$, $(1,1/4)$, $(1/4,1/4)$, $(1/4,1)$, and $(1/4,4)$ are plotted in green, yellow, cyan, blue, and black.}
    \red{(d) Map of (i) multi-polarized ($n=2$) domain wall speed under $k_1=10^{-3}$ and (ii) polarization transition width under $k_1=0$.}
    \red{Red triangle encloses $(f_v,f_\varphi)$ conducive to pinned domain walls.}
    Parameter values: \red{(a,b)} identical to Fig. \ref{fig:Model} caption, except $k_1=2\times10^{-3}$.}
	\label{fig:VDW_Motion}
\end{figure*}

The availability of multiple transition pathways affords the metamaterial an additional response. 
While the domain wall is topologically protected by the equilibrium states, $\mathbf{s}_{1,2}$, the polarization is not; therefore, it is possible for one polarization to be transformed into the other without violating the \red{topological} constraints.
To demonstrate, we modify the unit cell so that the local potential exhibits degenerate ground states, $\mathbf{s}_{1,2}$, and a central energy barrier comparable to \red{(but still greater than)} that of the \red{low-barrier} pathways.
In addition, we set $\varphi_0<0$ so that, of the two pathways, the transition via P1 is the more energetically expensive.
Within the resulting metamaterial, we initialize a static domain wall of type P1.
Subsequently, following the earlier procedure, for the lattice sites of cross-section, $j=0$, we prescribe a transformation from P1 to P2, which generates two, counter-propagating transition waves whose motion switches the polarization of the supporting domain wall (MOV. S3) and is driven by the potential difference between domain wall cross-sections of type P1 and P2, \red{respectively}.
The wavefronts are marked by a point of zero polarization -- denoted, ``switch point" -- where one \red{polarization} gives way to the other.
Through a series of snapshots, Fig. \ref{fig:Transformations}d depicts the motion of the two polarization-switching transition waves; \red{consequently, until the transformation process is complete,} the supporting domain wall exhibits more than one polarization simultaneously.
Figure \ref{fig:Transformations}e further reveals the process by which the wave effectuates the transformation, showing unit cell configurations traverse the local energy landscape encompassed by paths, P1 and P2, including the central barrier (MOV. S3).

These results are the first metamaterial examples of interfacial transition waves, which propagate by transforming \red{a configurational} boundary.
Given that both the interface-eliminating and polarization-switching transitions require the support of and, thus, are subordinate to a pre-existing transition structure (e.g., domain wall, \red{polarization interface}), these \red{are} \red{secondary} transitions \red{within a hierarchy of the same}.
Similar to a current-carrying wire, the configurational boundary supporting interfacial transition waves functions as a conductive pathway within an insulating medium.
With further development \red{to manipulate configurational boundaries (e.g., to realize arbitrary, stable, and re-positionable paths within the supporting metamaterial),} configurational boundaries may form the basis of reconfigurable mechanical circuits supporting interfacial transition wave signals.

\subsection{Domain Wall Retardation and Self-Pinning}
The propagation of polarization-switching transition waves causes the supporting domain wall to exhibit multiple polarizations simultaneously, i.e., to be \red{multi}-polarized.
In the previous section, the supporting domain wall is static; however, here, the situation is reversed as we investigate the effect of a static, spatially non-uniform polarization on domain wall motion.
To this end, \red{unit cells have}
$\Delta\psi>0$ and $\varphi_0=0$, \red{respectively, to support a potential driving and} to eliminate the potential difference between domain wall cross-sections of type P1 and P2.
Subsequently, we stimulate transition wavefronts with spatially varying polarization within a $52\times101$ metamaterial with periodic boundary conditions in the $j$-dimension: unit cells along $i=-50$ undergo prescribed state transformations via either the P0, P1, or P2 pathways as appropriate to manifest a periodic distribution of polarization in the $j$-dimension.
Regardless of the local transformation pathway, to stimulate a planar wavefront in the $v$ component \red{of the order parameter}, the excitation ensures a uniform $v$ along the excitation boundary.

Figure \ref{fig:VDW_Motion}a (MOV. S4) presents three cases of domain wall motion with spatially periodic polarization, featuring zero, two, and four switch points.
Comparing the positions of the wavefronts after \red{equal} propagation time, it is apparent that increasing \red{the number of switch points}, $n$, correlates with a reduction in the propagation speed.
Furthermore, \red{for each scenario}, examining \red{the} domain wall contour reveals that, despite stimulating planar wavefronts in the $v$ component, in time, wavefronts of non-uniform polarization become distorted: domain wall cross-sections nearer switch points lag those at greater distance.
Repeating the study within a $N\times101$ metamaterial array, we investigate \red{the dependence of the domain wall speed on the density of switch points, which goes as $n/N$.}
\red{The results in Fig. \ref{fig:VDW_Motion}b show the wave speed and switch point density to be inversely related: for $n>0$, the wave speed increases with $N$;
although, the speed is bounded above by that of a transition wave of homogeneous polarization.}


\begin{figure*}[t]
    \centering
    \includegraphics{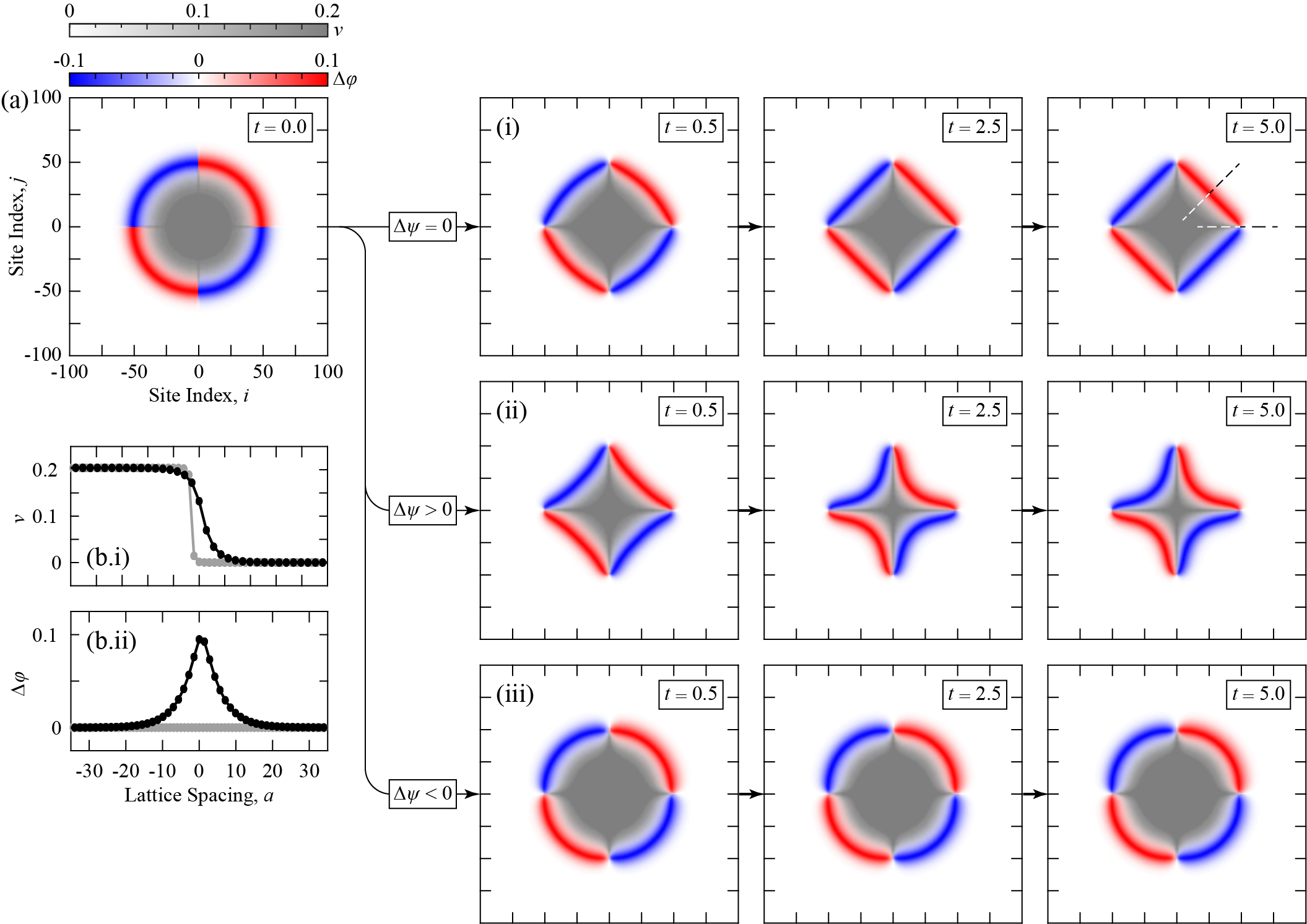}
	\caption{Domain Patterning.
    (a) Snapshots (at $t\times10^{-3}$) depicting the evolution of an initially circular domain wall with four switch points evenly distributed along the perimeter under (i) $\Delta\psi=0$, (ii) $\Delta\psi>0$, and (iii) $\Delta\psi<0$.
    With switch points as the pinning mechanism, domains subject to $\Delta\psi\geq0$ develop polygonal equilibrium morphologies to minimize the potential.
    (b) Cross-sections of the domain wall depicted in (a.i) at $t=5\times10^3$ through a switch point (i.e., along $j=0$) (grey) and away from switch points (i.e., along $j=i$) (black).
    Parameter values: identical to Fig. \ref{fig:Model} caption, except $(k_v,k_\varphi)=(1/3,5)k$, and (a.i) $k_1=0$ and (a.ii) $k_1=1.5\times10^{-3}$ and $v=-1/5$.}
	\label{fig:Pinning}
\end{figure*}

The above results are attributable to the discrete nature of the host medium.
Each advancement of the domain wall by one lattice spacing requires the reconfiguration of the \red{constituent} unit cells, which adjusts the total potential.
The difference \red{between} the maximum and minimum potential values during this process constitutes a translation energy barrier analogous to the Peierls-Nabarro barrier in dislocation theory\cite{PeierlsPPS1940,NabarroPPS1947,PeyrardPD1984,KivsharPRE1993,KevrekidisPD2000}.
As kinetic energy is converted to overcome the barrier \red{(which shares the spatial periodicity of the underlying lattice), the total kinetic energy oscillates (Fig. S2a) and} the wave mean velocity decreases.
An increase in the coupling strength, $k$, \red{decreases the variation between neighboring unit cells within a domain wall cross-section and, consequently}, the degree of reconfiguration required of unit cells per increment of wave advancement; \red{thus, reducing} the \red{translation} energy barrier.
Consider two cross-sections extracted from a domain wall of non-uniform polarization (Fig. S2b).
Relative to its counterpart, the profile extracted further from (nearer to) a switch point is more diffuse (localized) and, thus, the corresponding cross-section is less (more) impacted by discreteness effects, including reduced propagation speed.
As cross-sections nearer switch points propagate more slowly, the wavefront becomes distorted; moreover, to accommodate slower cross-sections, the propagation speed of the whole wavefront is reduced.
Thus, through enhanced discreteness effects at switch points, multi-polarization induces a retarding effect on domain wall motion.

\red{Inspired} by the above understanding, we \red{consider the possibility of exacerbating} the retarding effect of switch points toward preventing wave propagation entirely, i.e., realizing a self-pinning domain wall, which is arrested by an internal (and reversible) \red{means} -- localized, enhanced discreteness effects -- rather than by an \red{(potentially permanent)} external mechanism (e.g., lattice defect\cite{FrazierAM2017,JinPNAS2020}).
To this end, \red{given the importance of the coupling strength to the translation energy barrier and, therefore, the severity of discreteness effects}, $k_v$ and $k_\varphi$, which hitherto have been of equal strength, $k$, are made disparate.
\red{For different values of the inter-site coupling pair, $(k_v,k_\varphi)$, Fig. \ref{fig:VDW_Motion}c plots the steady-state propagation speed of an initially static multi-polarized domain wall ($n=2$) under a range of potential driving $\Delta\psi$ mediated by $k_1$.}
\red{Predictably, the propagation speed increases with the driving; however, for a given $\Delta\psi$, the speed is diminished in circumstances where $k_v<k_\varphi$.}
\red{Indeed, for certain $k_v<k_\varphi$ and $\Delta\psi$ below a threshold value, the speed is zero, i.e., the otherwise mobile domain wall is pinned by the switch point.}
\red{For a more comprehensive perspective, a sweep of $(k_v,k_\varphi)$ is undertaken for a fixed potential driving and the results presented in Fig. \ref{fig:VDW_Motion}d.}
\red{In particular, Fig. \ref{fig:VDW_Motion}d.i maps $(k_v,k_\varphi)$ to the domain wall speed, revealing the conditions for which self-pinning occurs.}
\red{Consistent with the earlier observation, the results suggest that the combination of a weak $k_v$ and a strong $k_\varphi$ is conducive to self-pinning.}
\red{Over the same parameter sweep, we also record the width, $W$, of the polarization transition region about the switch point ($\Delta\psi=0$ to prevent domain wall distortion), which correlates with the number of cross-sections experiencing enhanced discreteness and, therefore, contributing to domain wall retardation.}
\red{To this end, we fit}
\red{\begin{equation*}
    \varphi(\bar{y})=\frac{\varphi_\mathrm{max}-\varphi_\mathrm{min}}{2}\tanh{\left(\frac{a\bar{y}}{W}\right)},
\end{equation*}}
\hspace{-3.5pt}\red{to the $\varphi$-component of the domain wall cross-sections, where the normalized wave width, $W/a$, is a tuning coefficient\cite{Dauxois2006}; $\varphi_\mathrm{min}=\mathrm{min}(\varphi_{ij})$ and $\varphi_\mathrm{max}=\mathrm{max}(\varphi_{ij})$, and $\bar{y}$ is a dimensionless coordinate parallel to the $j$-dimension.}
\red{Figure \ref{fig:VDW_Motion}d.ii maps $(k_v,k_\varphi)$ to the width of the polarization transition region.}
\red{Comparing the maps in Fig. \ref{fig:VDW_Motion}d reveals that the $(k_v,k_\varphi)$ conducive to self-pinning also correlate with broader polarization transition widths.}
\red{Therefore, we conclude that broadening the width of the polarization transition region about a switch point encourages domain wall pinning.}
\red{Apparently}, a weaker $k_v$ enhances discreteness effects, especially in the vicinity of switch points.
A stronger $k_\varphi$ expands the region (i.e., number of domain wall cross-sections) about a switch point that \red{experiences} enhanced discreteness effects, which increases the per switch point retarding capacity \red{and the prospects for self-pinning under a given potential driving}.

\subsection{Complex Domain Patterning}
Beyond its novelty in the metamaterial literature, self-pinning may have practical relevance in stabilizing complex (i.e., non-laminar) domain morphologies \red{toward}, e.g., domain patterning in 2D systems \red{in support of morphable surfaces impacting fluid flow and, in leveraging the distinct physical properties of domains, tunable material response\cite{RamakrishnanJAP2020,RamakrishnanJASA2023}}.

To \red{investigate the efficacy of self-pinning for stabilizing complex 2D domain morphologies}, within a $201\times201$ metamaterial array with periodic boundary conditions, we consider the fate of initially circular domains of $\mathbf{s}_2$ -- as defined by the $v$ contours -- \red{within} a background of $\mathbf{s}_1$ with, respectively, zero, two, and four switch points periodically-distributed along the perimeter.
Curved domain walls are \red{inherently} unstable: \red{curvature increases the interfacial potential energy}, which the system resolves by minimizing the \red{interfacial length}.
Accordingly, absent local driving, the circular domain with homogeneous boundary polarization symmetrically contracts and, ultimately, vanishes (MOV. S5).
To encourage the self-pinning of the \red{multi}-polarized domain walls, following the above rationale, we set $(k_v,k_\varphi)=(1/3,5)k$.
In addition, for initial studies, $k_1=0$ to void of any local driver of domain wall motion (i.e., $\Delta\psi=0$).

For initially circular domains of radius, $r=50a$, with periodic polarization, the portions of the domain wall sufficiently far from switch points adjust to minimize the local curvature; however, in the vicinity of a switch point, the domain wall is pinned (MOV. S5), the result of the locally insurmountable translation energy barrier.
Self-pinning precludes the complete collapse of the domain; instead, the initially circular domains develop new meta-stable, polygonal morphologies.
Figure \ref{fig:Pinning}a highlights the case of four switch points.
We analyze two domain wall cross-sections extracted from the \red{meta-stable} morphology (Fig. \ref{fig:Pinning}b): one through a switch point; the other through a point midway between two switch points.
In particular, we compare the \red{widths, $W$, of the transition regions}, which is a measure that inversely correlates with the severity of the discreteness effects \red{within} the cross-section.
\red{To this end, we fit}
\red{\begin{equation*}
    v(\bar{x})=\frac{v_{\mathbf{s}1}-v_{\mathbf{s}2}}{2}\tanh{\left(\frac{a\bar{x}}{W}\right)}-\frac{v_{\mathbf{s}1}-v_{\mathbf{s}2}}{2},
\end{equation*}}
\hspace{-3.0pt}to the $v$-component of the domain wall cross-sections (Fig. \ref{fig:Pinning}b.i).
For the cross-sections through and away from the switch point, respectively, we find \red{$W/a\approx0.391$} and \red{$W/a\approx2.668$}, indicating the switch point cross-section to be most impacted and suggesting enhanced discreteness effects as responsible for the pinning behavior.

We, \red{now}, adjust the metamaterial microstructure to \red{investigate} the response under \red{local driving (i.e., $\Delta\psi\neq0$)}.
Remarkably, for modest $\Delta\psi\neq0$, \red{switch points are still able to} stabilize a complex domain morphology \red{via discreteness-induced self-pinning}; moreover, one with curved boundaries.
For $\Delta\psi>0$, \red{the domain boundary bows inward as the driving tends to collapse the originally circular domain, but is resisted by self-pinning at the switch points.}
\red{Setting} $\Delta\psi<0$ \red{has the opposite effect as, way from the pinned switch points}, the domain boundary \red{bows} outward until the interfacial forces balance the local driving forces.

\red{In attempting} to stabilize domains \red{via the self-pinning mechanism}, the lattice density in the direction of the \red{local} domain wall motion \red{is} a key factor: switch points are most effective \red{in pinning those cross-sections with more participating unit cells}.
For the square lattice, \red{the arrangement of unit cells along} the $i$ and $j$ dimensions \red{is} the most dense.
In addition, if the \red{axis containing the switch points is not aligned with the direction of domain wall motion (i.e., local normal to the $v$ contour)}, then the switch points will migrate along the domain wall in order to minimize the interface energy (MOV. S6).
The direction of the migration is toward the immediate portion of the domain wall with the smaller local curvature.

\section{Summary}
In summary, this article introduces a class of bi-stable metamaterial whose microstructure, nevertheless, permits multiple transition pathways between the stable equilibrium states.
\red{This unique condition, the result of unconventional symmetry breaking, simultaneously, introduces the notion of polarization to the process of switching between states}.
Utilizing a lumped parameter model, we numerically examined the consequences of this condition for the transformation dynamics, revealing behavior hitherto unreported in the metamaterial literature.
We \red{are able to stimulate} two types of interfacial transition waves, whose propagation is confined to the narrow configurational boundaries formed by pre-existing domain walls and \red{which} may be consequential in the development of reconfigurable mechanical circuits.
In addition, we \red{are able to realize} domain walls of non-uniform polarization, \red{which} exhibit a reduced propagation speed due to an increase the translation energy barrier (a discreteness effect) \red{about points} where the domain wall switches polarization.
Given that \red{a switch point} is not \red{a} topologically protected \red{object}, switch points may be added/removed \red{as desired} and, thus, represent an approach to reversibly tune the domain wall speed without altering the underlying microstructure.
\red{Moreover}, below a threshold potential driving, \red{a} switch point \red{functions} as an intrinsic, non-structural pinning mechanism to prohibit domain wall motion.
\red{To demonstrate} the pinning ability, we stabilize 2D domains whose boundaries feature multiple switch points, a result relevant to, e.g., reversible domain pattering for tunable materials and morphable structures/surfaces.
These results illustrate some of the unique consequences of multiple transition pathways in bi-stable metamaterials and motivate \red{further} study of multi-stable metamaterials with unconventional transition processes.



\bibliographystyle{ieeetr}
\bibliography{xRefs_MJF}
\end{document}